\documentclass[conference]{IEEEtran}

\usepackage[utf8]{inputenc}
\usepackage{amsmath}
\usepackage{graphicx}
\usepackage{xcolor}
\usepackage{tikz}
\usepackage{pgfplots}



\makeatother

\usepackage{stfloats}

\begin{document}
\title{PPContactTracing: A Privacy-Preserving Contact Tracing Protocol for COVID-19 Pandemic}

\author{\IEEEauthorblockN{Priyanka Singh\IEEEauthorrefmark{1},
Abhishek Singh\IEEEauthorrefmark{2},
Gabriel Cojocaru\IEEEauthorrefmark{2}, 
Praneeth Vepakomma \IEEEauthorrefmark{2},
and
Ramesh Raskar\IEEEauthorrefmark{2}}
\IEEEauthorblockA{\IEEEauthorrefmark{1}Dhirubhai Ambani Institute of Information and Communication Technology\\
Gandhinagar, Gujarat, India\\ Email: Priyanka\_Singh@daiict.ac.in}
\IEEEauthorblockA{\IEEEauthorrefmark{2}MIT Media Lab\\
Cambridge, MA, USA\\
Email: \{abhi24,r\_eality,vepakom,raskar\}@mit.edu}}

\maketitle

\begin{abstract}
Several contact tracing solutions have been proposed
and implemented all around the globe to combat the spread
of COVID-19 pandemic. But, most of these solutions endanger
the privacy rights of the individuals and hinder their widespread adoption. We propose a privacy-preserving
contact tracing protocol for the efficient tracing of the spread of the
global pandemic. It is based on the private set intersection (PSI)
protocol and utilizes the homomorphic properties to preserve
the privacy at the individual level. A hierarchical model for the representation
of landscapes and rate-limiting factor on the number of queries have been adopted to maintain the efficiency of the protocol.
 
\end{abstract}

\IEEEpeerreviewmaketitle

\section{Introduction}

The pandemic caused due to novel COVID-19 has influenced the daily life of almost every individual. In the face of such a situation, the researchers across the entire globe are working towards a solution of to cater to this problem. Many systems have been designed to identify the disease carriers, trace the probable points of contact and to notify such individuals. This process is referred to as contact tracing and can be very labor intensive, time-consuming, memory constrained and privacy intrusive. However, it is very crucial for government and health officials to contain the outbreak of such a pandemic. 

Many contact tracing apps have been proposed since the spread of COVID-19 but privacy preservation has been a major bottleneck. The Singapore government built the TraceTogether app for smartphones that used the low energy of Bluetooth to periodically generate a random identifier called as `contact event number (CEN)'\cite{Government}.  This number is broadcasted to other mobile users in the vicinity of the device and the receivers keep recording a list of the CENs. Later on, once a person is diagnosed as COVID-19 positive, the respective CEN is shared with the public and it is matched against the CENs recorded at every device and this helps determine the susceptibility of any user. Here, a central database was maintained by the government where they saved a record of the CENs and the associated phone numbers and identities. Hence, this model assures privacy of an individual from other app users but assumes a complete trust over the central authority. This can be seen as a breach of privacy preserving principles when multiple parties have to trust a centralized authority in a distributed setting.
 
The Australian government came up with CovidSafe, a contact tracing app based on Bluetooth technology. It requires an initial registration and then keeps on collecting information about people in close proximity for 21 days. If anyone is diagnosed as COVID-19 infected, then that person shares the stored information with the state health officials so that they can alert the individuals that may have been in close contact. This app received criticism because of the potential for massive privacy loss due to information being held by the government.
       
        Some other models that put more trust at the users end rather than on a central authority are COVID Watch \cite{Rhys} and CoEpi \cite{Dana}. In here, the recorded database is made available to all the users of the application and it may be prone to privacy attacks from their peers \cite{cho2020contact}. The main bottleneck with the Bluetooth based technologies in general is that these signals can be obstructed by buildings, walls or objects and intervene with the information of probable points of contact. In addition, these solutions are efficient and feasible in settings where the devices are within close proximity. Additionally, a constraining demerit of Bluetooth based technologies is that they can just detect person-to-person contact but fail in detecting transmission through surfaces in contact.  
 
 GPS based technologies that record location histories take care of this issue better as they can compensate for such missing chunks of data based on recorded timestamps. They are also based on a ubiquitously available technology that most of the digital devices already capture. This helps to cater to a broader base of people and it hence can be incorporated with larger potential impact. India launched Aarogya Setu app based on a combined technology of GPS and Bluetooth in order to track the COVID-19 infected individuals. Through Bluetooth, the app can track people within a range of six feet and check for the infected individuals by scanning through the database of known infected individuals. Location history was used to confirm whether the location lies in the infected areas or not. That said this app as well shares some information of the registered individuals with the government and health officials and treats them as a centralized trusted authority.
%

\par In the United States, there has been no nationwide contact tracing app although companies like Google and Apple launched Exposure Notification (GAEN) APIs to support contact tracing which some states like Virginia, South Carolina, North Dakota, Alabama agreed to adopt. Research groups at MIT Media Lab discussed about the risks and opportunities involved in adoption of such technologies for the society. They proposed an open source, decentralized solution such as MIT Private Kit: Safe Paths (now PathCheck) to contain the spread of the pandemic. This allows for both GAEN as well as GPS and location based digital contact tracing.\\
In this paper, we propose a privacy preserving solution for contact tracing that preserves the privacy of the infected as well as the healthy person and helps to track information about the possible spread of the pandemic.\\
In terms of existing solutions, `private set intersection' (PSI) has a potential to address the privacy concerns detailed above. Meadows \cite{meadows1986more} proposed the first PSI protocol in 1986. This protocol was based on Diffie-Hellman (DH) key exchange protocol, which is a seminal work. This used the commutative property of DH function securely to compare and match the credentials of two parties. An example using this protocol for COVID-19 applications can be seen in \cite{berke2020assessing}. In 2004, Freedman \cite{freedman2004efficient} presented two efficient protocols: one in a semi-honest adversary model and another in a malicious adversary model. These models were based on homomorphic public key cryptosystem (PKC), balanced hashing, and polynomial interpolation. These were shown to be secure under the standard model and random oracle model. In 2015, a proposal came from Debnath and Dutta \cite{debnath2015secure} based on multiplicative homomorphic public key cryptosystem and bloom filters \cite{bloom1970space}. In 2016, a formal simulation-based security proof under the malicious adversaries model was proposed by the extended approach of \cite{freedman2004efficient}.
We propose a new privacy preserving solution for contact tracing based on PSI with several advantages as detailed below. The main contributions of the proposed approach is that it ensures the privacy of healthy and the infected person as only encrypted data is shared based on a client/server model with the healthy person/infected person. Most of the computations in our setting are done at the server end without it knowing anything in the whole process as the computations are done on top of encrypted data. This resolves the trust issue with regards to a centralized authority as detailed in existing apps and solutions referred to above. Another big advantage of the proposed scheme is in terms of efficiency as we employ a hierarchical structure to represent the several landscapes that depend on a contact tracing solution. In addition, we place a restriction  on the number of queries from the client-side in order to further secures any leakage of any infected individual information maintained at the server end.\\
\section{Preliminaries}
\subsection{Paillier Cryptosystem}
\label{sec:paillier}

The Paillier cryptosystem is an additive, partially homomorphic, and asymmetric encryption scheme with public key and private key pair as ($n$ and $g$) and ($\lambda$ and $\mu$) respectively ~\cite{paillier1999public}. 

The encryption of a plaintext message $m$ with public key ($n$, $g$) is done as follows:
\begin{eqnarray}
c ~=~ E(m,r;g,n) ~=~ \mbox{mod}(g^m \times r^{n}, n^{2}),
\label{eqn:encrypt}
\end{eqnarray}
where $r$ is a random integer satisfying $0<r<n$. The incorporation of this random value ensures that the same plaintext is encoded as different ciphertexts under the same public key.

A ciphertext message $c$ is decrypted as follows:
\begin{eqnarray}
D(c;\lambda,\mu,n) & = & \mbox{mod} \left( \frac{\mbox{mod}(c^\lambda,n^2) - 1}{n} \times \mu, n \right).
\end{eqnarray}
It supports the following homomorphic properties: 
\begin{itemize}
	\item  The sum of two plaintexts $m_1$ and $m_2$ is equivalent to the decrypted product of corresponding ciphertexts $c_1$ and $c_2$ : 
	\begin{eqnarray}	
    \mbox{mod}(m_1+m_2,n) =	D(\mbox{mod}(c_1 \times c_2, n^2);\lambda, \mu, n)
	\label{eqn:homomorphic1}
	\end{eqnarray}
	\item  The product of a scalar $s$ with a plaintext $m_1$ is equivalent to the decrypted exponentiation of the corresponding ciphertext $c_1$ with the scalar:
	\begin{eqnarray}	
    \mbox{mod}(s \times m_1,n) = D(\mbox{mod}(c_1^s, n^2);\lambda, \mu, n) 
	\label{eqn:homomorphic2}
	\end{eqnarray}
\end{itemize}

\subsection{Threat Model}
The threat model is constructed for two sets of parties; the healthy person and infected person. For the healthy person, we consider a semi-honest adversary model. In the semi-honest adversary model, the client performs their computation honestly but tries to learn maximum information from the data they receive. For the infected person, in the protocol described in section~\ref{sec:protocol} we assume that the infected person trusts the server with their data which is true in the case of a contact tracing scenario where the infected person gets interviewed by the contact tracers at the healthcare authority. In addition, we have discussed an extension in the section~\ref{sec:untrusted_server} where the infected party also maintains a zero-trust model like the healthy person.

\subsection{Hierarchical Partitioning of Landscape}
\begin{figure}[h!]
	\centering
	\includegraphics[scale=0.65]{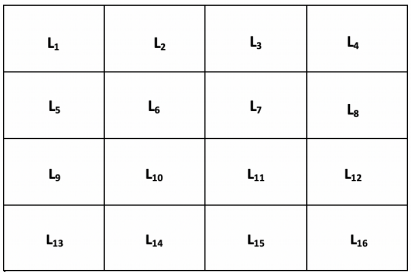}
	\caption{A Portion of the landscape grid where each cell in the grid represents a particular position at a given time.}
	\label{fig:Grid}
\end{figure}

Partitioning a geographic space is a pretty standard terminology and can be done either using geohashes for square shaped grid cells or hexagonal  using the H3 grid which is a global geo-spatial indexing system. An example of the square shaped grid is shown in Figure~\ref{fig:Grid}. This 3-dimensional grid is made accessible to all the users.

The proposed protocol takes as input, the geographic location (GPS coordinates of latitude and longitude) along with the time stamp of the location as the third dimension and maps this spatio-temporal location to discrete point-intervals. There lies a trade-off between the precise tracing of individuals and the amount of data that needs to be processed. If the time interval is made very small say 30 seconds, it will give more finer details of contact contact tracing compared to a time interval of 5 minutes but definitely with a significant increase in the bulk of data storage and computational resource requirements. The appropriateness of time interval and actual partitioning scheme can be decided based on the experiments and implementation details.


\section{The Proposed Protocol}
\label{sec:protocol}
\begin{figure}[h!]
	\centering
	\includegraphics[scale=0.65]{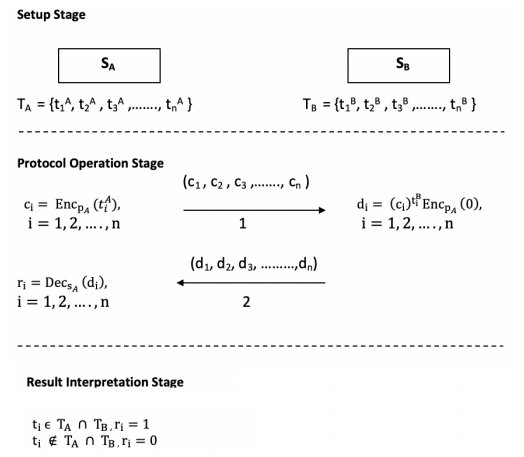}
	\caption{Overview of the Proposed Protocol}
	\label{fig:ProtocolDiag}
\end{figure}

We present a privacy preserving protocol that is useful for contact tracing. It helps in detection of any matches for a user with the diagnosed  infected users based on sharing the GPS trajectories. It operates as a client-server model where the server maintains the database of infected people and a normal user is treated as a client. An overview of the proposed protocol is shown in Figure~\ref{fig:ProtocolDiag}.

The user record their own GPS . The user shares his encrypted GPS history with the server and the server then operates on top of this encrypted data and sends back the processed encrypted data or encrypted result back to the user. The user decrypts the encrypted information sent to him and deduces based on the decrypted value whether he had any common visited area or a probability of getting suspected. There is no privacy leakage in the entire protocol as both the client and server are exchanging encrypted information. The encryption is done based on PKI  based public/secret key pair for each entity. The server learns nothing about the co-ordinates of the user whereas the user gets to know the intersection result. 

\subsection{Setup Stage}
Let us suppose the landscape $L$ is divided into square grid cells  ${l_{1},l_{2},l_{3},\ldots l_{n}}$ in order.
Based on the client-server model, the user can be considered as the client $S_{A}$ and the diagnosed infected user as the server $S_{B}$. The locations visited by the entity can be represented by a bit vector of same length as the number of square grid cells. If a grid cell $l_{i}$ has been recently visited, it is represented by a corresponding bit value of $1$ else the replaced with a $0$ value maintaining the same order as  of the original grid cells. Let us represent the bit vectors for the client $S_{A}$  and the server $S_{B}$ by $\overrightarrow{{t_{i}}^{A}}$  and $\overrightarrow{{t_{i}}^{B}}$ respectively. Each entity encrypts every bit of their trajectory using a public/private key pair. In the client-server model, let $(p_{A}, s_{A})$ and $(p_{B}, s_{B})$ be the corresponding keys. The encryption and decryption are represented as functions: $Enc_{p_{k}}()$ and $Dec_{s_{k}}()$ using the public key/secret key pair as $(p_{k}, s_{k})$ respectively.

\subsection{Set Operation Stage}
The client-server interacts in the following manner to obtain the set intersection.

1) The client $S_{A}$  encrypts each bit of its trajectory bit vector ${t_{i}}^{A}$ using its public key $p_{A}$

\begin{equation}
c_{i} = Enc_{p_{A}}({t_{i}}^{A})
\end{equation}

This ciphertext $c_{i}$ $(1 \leq i \leq n)$ is then sent to the server.

2) The server receives the encrypted ciphertext $c_{i}$ $(1 \leq i \leq n)$ and processes on top of this encrypted ciphertext to obtain 
\begin{equation}
d_{i} = {(c_{i})}^{t_{i}^{B}}. Enc_{p_{A}}(0)
\label{eqn:server_compute}
\end{equation}

This processed ciphertext $d_{i}$ $(1 \leq i \leq n)$ is then sent to the client.

3) The client decrypts the received ciphertext $d_{i}$ using its secret key $s_{A}$
\begin{equation}
t_{i} = Dec_{s_{A}}({d_{i}})
\end{equation}

These $t_{i}'s$ fetches the information of the set intersection or possible matches based on following:

\begin{equation}
t_{i} = \left\{\begin{matrix}
1 & if \  l_{i} \in S_{A} \bigcap S_{B} \\ 
 0 &  if \  l_{i} \not\in  S_{A} \bigcap S_{B}
\end{matrix} \right.
\end{equation}

\subsection{Result Interpretation Stage}

 The proposed protocol preserves the privacy of both client and server.  The client shares its encrypted trajectory bit vector $c_{i}$ to the server. The server processes the ciphertext and sends the processed ciphertext $d_{i}$  to the client. The client then decrypts the received processed ciphertext to obtain information about the matching or overlapping regions based on decrypted bits $t_{i}$. Positions where $t_{i}$'s are $1$ represent the overlapping or matched locations else they have no suspicion of contact.
   The privacy of the server is preserved by multiplication of the ciphertext by encrypted zero after exponentiation with the trajectory bit vector $t_{i}^{B}$ of the server in step 2 of the set operation phase. 
 This operation is exploiting the additive homomorphic property that allows exponentiation and multiplication on the ciphertexts equivalent to the scalar multiplication and addition in the corresponding plaintexts.
 \begin{equation}
\begin{matrix}
 d_{i}  & = & {(c_{i})}^{t_{i}^{B}}. Enc_{p_{A}}(0)\\ 
          &  =  &  \ Enc_{p_{A}}( t_{i}^{A} \times t_{i}^{B} + 0) \\ 
         &   =  &  Enc_{p_{A}}( t_{i}^{A} \times t_{i}^{B}) \\ 
\end{matrix}
\end{equation}
 
Multiplication of encrypted zero is equivalent to addition of zero in the plaintext. Thus, it makes no difference to the result but helps to preserve the privacy of the server. If this multiplication by encrypted zero is not done in the processed ciphertext, then it may leak the trajectory bit vector $t_{i}^{B}$  of the server. 

\begin{equation}
d_{i} = \left\{\begin{matrix}
c_{i} & if  \ \  t_{i}^{B} = 1  \\ 
 1 &  if \  \ t_{i}^{B} = 0  
\end{matrix} \right.
\end{equation}
 
 \section{Modified protocol for cardinality of set intersection}
For applications that aims to preserve more privacy to the parties interacting and does not want to reveal anything apart from the cardinality of the set intersection. Suppose two parties $S_{A}$ and $S_{B}$ possess sets $T_{A}$ and $T_{B}$ and interact to compute the set intersection function $f(T_{A}, T_{B})$  = $\left ( \perp , \left | T_{A} \bigcap T_{B} \right | \right )$, $S_{A}$ gets to know only the cardinality of the set intersection without any other information and $S_{B}$ learns nothing.  The details are as follows:

\subsection{Setup Stage}
The setup is same as the proposed protocol. The bit vectors $\overrightarrow{{t_{i}}^{A}}$  and 
$\overrightarrow{{t_{i}}^{B}}$ represent the client $S_{A}$  and the server $S_{B}$. $(p_{A}, s_{A})$ and $(p_{B}, s_{B})$ as the (public key, secret key) pairs. 

\subsection{Set Operation Stage}
The client-server interacts in the following manner to obtain the set intersection.

1) The client $S_{A}$  encrypts each bit of its trajectory bit vector ${t_{i}}^{A}$ using its public key $p_{A}$

\begin{equation}
c_{i} = Enc_{p_{A}}({t_{i}}^{A})
\end{equation}

This ciphertext $c_{i}$ $(1 \leq i \leq n)$ is then sent to the server.

2) The server receives the encrypted ciphertext $c_{i}$ $(1 \leq i \leq n)$ and processes on top of this encrypted ciphertext to obtain 
\begin{equation}
d = \prod_{i}^{n} {(c_{i})}^{t_{i}^{B}}. Enc_{p_{A}}(0)
\end{equation}

This processed ciphertext $d$ is then sent to the client.

3) The client decrypts the received ciphertext $d$ using its secret key $s_{A}$
\begin{equation}
t = Dec_{s_{A}}({d})
\end{equation}

This $t$ gives the information about the cardinality of the  set intersection.  

\subsection{Result Interpretation Stage}
   
 The proposed protocol preserves  more privacy of both client and server.  The client gets to know the cardinality of the set intersection whereas the server does not get any information in the whole process.
 
The client encrypts its trajectory bit vector and sends  the  encrypted trajectory bit vector $c_{i}$ to the server. The server processes the ciphertext and sends the processed ciphertext $d$  to the client. The client decrypts $d$  and obtains $t$ which gives the information about the cardinality of the set intersection or the matching regions.  
  
The processing at the server in step 2 exploits the additive homomorphic property.  
 \begin{equation}
\begin{matrix}
 d_{i}  & = & { \prod_{i}^{n}(c_{i})}^{t_{i}^{B}}. Enc_{p_{A}}(0)\\ 
          &  =  &  \prod_{i}^{n} ((Enc_{p_{A}}( t_{i}^{A}))^{t_{i}^{B}}). Enc_{p_{A}}(0) \\ 
          &  =  &  \prod_{i}^{n} (Enc_{p_{A}}(t_{i}^{A} \times t_{i}^{B})). Enc_{p_{A}}(0) \\ 
         &   =  &  Enc_{p_{A}}(\sum_{i}^{n}( t_{i}^{A} \times t_{i}^{B})). Enc_{p_{A}}(0) \\ 
         &   =  &  Enc_{p_{A}}(\sum_{i}^{n}( t_{i}^{A} \times t_{i}^{B}) + 0) \\ 
         &   =  &  Enc_{p_{A}}(\sum_{i}^{n}( t_{i}^{A} \times t_{i}^{B})) \\ 
\end{matrix}
\end{equation}
 
Thus, the privacy of the server is preserved in step 2 of the set operation phase by multiplication of the ciphertext by encrypted zero after exponentiation with the trajectory bit vector $t_{i}^{B}$. Nothing is leaked to the server and the client only gets to know the cardinality of the intersection of the sets after decryption of the processed cipher received from the server.  

\begin{equation}
\begin{matrix}
   t      & =  & Dec_{s_{A}}({d}) \\
          &  =  &  Dec_{s_{A}}(Enc_{p_{A}}(\sum_{i}^{n}( t_{i}^{A} \times t_{i}^{B}))) \\ 
          &  =  &  \sum_{i}^{n}( t_{i}^{A} \times t_{i}^{B})) \\         
\end{matrix}
\end{equation}
  
\section{Experiments and Analysis}

We perform all the experiments in Go. The only parallelization we perform is done through go-routines which creates a multi-threading environment. However, there are much more parallelization and system level optimization which we consider as part of the future work. The experiments were performed on a dockerized container with \textit{Intel Intel(R) Xeon(R) CPU E5-2650 v4 @ 2.20GHz, x86\_64} and \textit{470GB} of RAM. The experiment with varying set size is discussed in the table~\ref{tab:set_size} for private set cardinality and table~\ref{tab:set_size_cardinality} shows the result for overall computation when performing private set cardinality intersection. For our experiment with numbers in the same ballpark as the practical scenario of contact tracing, it takes roughly a minute or so for every client's computation on the server, as measured by wall clock time. The total computation time can be broken down into two parts - computation time by the server and time taken by a healthy person for the decryption of the results. Decryption of the results is in the order of milliseconds for the private set cardinality protocol because there is only one value which needs to be decrypted as the final answer.\\
One big advantage of this approach is that the server computation time stays constant even as we scale the number of infected population. This is due to the single vector held by the server for querying every healthy client

\begin{table}[]
    \centering
    \begin{tabular}{|c|c|c|c|c|}
         \hline
         Set Size & 512 bit & 1024 bit & 2048 bit & 4096 bit\\
         \hline
         $2^{10}$ & 347ms & 1.38s & 6.37s & 30.38s\\
         \hline
         $2^{11}$ & 700ms & 2.77s & 11.7s & 55.61s\\
         \hline
         $2^{12}$ & 1.16s & 4.42s & 20.2s & 1m41s\\
         \hline
         $2^{13}$ & 1.89s & 7.72s & 36.7s & 3m8s\\
         \hline
         $2^{14}$ & 3.2s  & 12.95s& 1m4s  & 5m50s\\
         \hline
         $2^{15}$ & 5.8s  & 24.89s& 2m4s  & 11m14s\\
         \hline
         $2^{16}$ & 11.73s& 48.82s& 4m2s  & 21m51s\\
         \hline
         $2^{17}$ & 22.62s& 1m38s & 7m56s & 43m12s\\
         \hline
         $2^{18}$ & 43s   & 3m48s & 15m37s& 2h14m\\
         \hline
    \end{tabular}
    \caption{Server computation time as reported by the wall clock time across different bit sizes of the key of the Pailier cryptosystem.}
    \label{tab:set_size}
\end{table}

\begin{table}[]
    \centering
    \begin{tabular}{|c|c|c|c|c|}
         \hline
         Set Size & 512 bit & 1024 bit & 2048 bit\\
         \hline
         $2^{10}$ & 3.35s & 12.11s & 1m6s\\
         \hline
         $2^{11}$ & 6.87s & 24.8s & 2m13s\\
         \hline
         $2^{12}$ & 13.07s & 51.25s & 4m28s\\
         \hline
         $2^{13}$ & 30.01s & 1m42s & 8m41s\\
         \hline
         $2^{14}$ & 59s  & 3m17s & 16m43s\\
         \hline
         $2^{15}$ & 1m55s & 6m11s & 31m27s\\
         \hline
         $2^{16}$ & 3m48s & 11m30s & 56m38s\\
         \hline
         $2^{17}$ & 7m3s & 21m5s & 1h44m\\
         \hline
         $2^{18}$ & 14m27s & 39m26s & 3h25m\\
         \hline
         $2^{19}$ & 25m54s & 1h11m & 6h39m\\
         \hline
    \end{tabular}
    \caption{Server computation time as reported by the wall clock time across different bit sizes of the key of the Pailier cryptosystem developed for private set cardinality protocol.}
    \label{tab:set_size_cardinality}
\end{table}

\subsection{Communication complexity analysis}
In our protocol, the client communicates only once a day since there is no need to query for exposure notification more than a day. Therefore, the communication complexity depends on how much data is transferred in a single batch transfer. Current GPS apps used for contact tracing record GPS data every five minutes, hence for simplicity, we also use the same number for our communication complexity analysis. Every five minutes of GPS scan leads to $288$ scans for any given day. Hence, user uploads $288 x N$ points in a single batch. Here, $N$ is the total size of the vector used to represent a single GPS scan which is also equal to the total number of cells in the location grid. We use one-hot representation of this vector with $M$ being the size of every element in the one-hot vector. Hence the communication complexity is $O(M*N)$ bits. From a practical standpoint, there is no significant communication overhead for individuals because they query the server only once a day. For the Server, it is possible to scale up receiving endpoint by simply using horizontal scaling or load balancing.\\
The communication complexity for the download is same as upload in the case of full private set intersection protocol. However, for the case of private set cardinality, the server only returns a single encrypted number to the client indicating whether there is a match or not.

\subsection{Computation complexity analysis}

\subsubsection{Exponentiation}

The bottleneck in the performance of encryption and decryption lies in the performance of exponentiation. In the case of encryption we need to raise $r$ to the power of $N$ and for decryption we need to raise $c$ to the power of $\lambda$ where $\log \lambda = O(\log N)$. The brute force approach of multiplying $2$ big integer takes $O(\log^2 N)$ but modern implementations such as that of Go's math/big store bits in big integers in chunks of size $W$ where $W$ is equal to the word size. \cite{GoMathBigMul} Assuming multiplication and addition over integers with $W$ bits takes constant time, the brute force complexity is reduced to $O((\frac{\log N}{W})^2)$. Upon a certain threshold of $\log_2(N)$ Karatsuba's $O((\frac{\log N}{W})^{1.58})$ algorithm becomes faster in practice. If we increase $log_2(N)$ to be big enough, we can use the Sch{\"o}nhage-Strassen $O(\log N \log \log N)$ algorithm for multiplication. There is a Java big integer implementation of Sch{\"o}nhage-Strassen which always uses it for integers bigger than $2^{2^{19}}$. \cite{JavaBigInteger} The exponentiation takes $O(E(N) \log N)$ per one exponentiation with either $E(N) = O(\frac{\log^2 N}{W^2})$, $E(N) = O(\frac{\log^{1.58}N}{W^2})$ or $E(N) = \frac{\log N \log \log N}{W^2}$ depending on how big is $\log_2(N)$. 

\subsubsection{Client}

The total complexity performed by the client on encryption is $O(E(N)(M\log N)$ and the same complexity is for decryption. In practice the encryption turns out to be twice slower on average than the decryption. One of the reasons behind this could be that on average we have $\log \lambda \approx \frac{\log N}{2}$.

\subsubsection{Server}

For the general PSI, we are performing $M$ additions, multiplications and encryptions of $0$, the costliest being the $0$ encryption. So the complexity is $O(E(N)(\log N)M$).

For the case when we only want to find the cardinality of the PSI, we perform only one encryption of $0$, so the complexity is $O(E(N)(\log N + M))$.

\subsubsection{Speeding up encryption}

The idea of choosing $r$ and taking it to the power $N$ is to have that $r^{N \lambda} \equiv 1 (\mod N^2)$ so that during the decryption it will vanish leaving only the relevant information. To achieve the same encryption, every time we want to encrypt a vector of messages, we will uniformly chose a random $r_0$. Then for every element $m$ in our vector, if we choose a number $x$ uniformly from interval $[0, \lambda - 1]$ we can encrypt $m$ as $mod(g^m \times r_0^{xN}, N^2)$. This way if we have a sequence of messages to encrypt, we will encrypt them independently of each other. If we choose $x$ to be uniformly from interval $[0, 2^K - 1]$ we will make the encrypted sequence of messages dependent of each other since we are using only $2^K$ possible values. The hardness to break the encrypted sequence is the same as before, i.e. the hardness of factorisation and the trade-off is that if an encrypted message plaintext value is leaked then the adversary will be able to find the whole sequence using $O(2^K)$ tries.. If we use this method we can get the encryption down to $O(E(N)(\log N + M))$.

\subsubsection{Choosing the security}

The number of bits used in $N$ is crucial to the security of protocol. Both in practice and theoretically we found that if we double the number of bits in $N$ then the algorithm gets slower by a factor ranging from $4$ to $8$. We recommend using $1024$ bits for full PSI and $4096$ bits + above-mentioned fast encryption for finding only the size of PSI. Upon increasing the number of bits too much, the protocol becomes unscalable. For example if the words size $W = 64$, then if we let $\log_2 N$ have $13$ bits $E(N) \log N$ reaches order of tens of millions, for $\log_2 N$ with $18$ bits, $E(N) \log N$ reaches order of hundreds of billions. Besides performance we would also require more space. For a sequence with $2^{17}$ elements and $4096$ bits security, we need $2^{31}$ bits to store only the encrypted elements.

\subsubsection{Future possible improvements}

To improve further on the current protocol we can use a different partially homomorphic encryption scheme that supports addition over ciphertexts. One such example would be Benaloh cryptosystem
which uses smaller exponents and thus reduces the time taken by the exponentiation. 

\section{Extension for different security constraints}
\subsection{Untrusted Server}
~\label{sec:untrusted_server}
In this setting, we assume that the central server can not be trusted even with regards to the data of infected person. We extend the current method as described in the section 4 by adding one extra server which solely performs the key exchange between the healthy and infected person database. Both healthy and infected party can download the keys from this key exchange server and encrypt their data and upload it for comparison on the computation server responsible for executing the operations mentioned in ~\ref{eqn:server_compute}.
\subsection{No upload of healhy user data to the server}
From the point of view of public trust as well as stringent security requirements, it is reasonable to put another constraint in the protocol by requiring no data upload by the healthy individual, be it as a plaintext or ciphertext. This can be achieved by having the computation to be performed on-device of the healthy person. However, direct sharing of encrypted or plaintext location information of the infected individual would allow the healthy person to perform brute-force attack and hence leak all of the information about the infected individual. This could be circumvented by having the healthy client know only the public key $p_A$ and hence they can only compute the encryption and the operations mentioned in the eq.~\ref{eqn:server_compute}. Then the results of the computation are shared with the server which decrypts the result and returns it back. This would prevent the brute-force attack because the server can enforce rate limiting to limit the queries which can be decrypted by a given healthy individual. If we want to hide the results of the computation from the server, the client can blind the outputs by multiplying with additional blinding factor, making the difficulty of recovering the results as difficult as the factorization problem. In principal, we achieve this extension by sending encrypted result of the computation instead of sending the encrypted plaintext. Note that this could hurt the efficiency of the protocol because now the whole computation load resides on the client device which is a handheld device. Hence, this solution comes with a trade-off between secrecy of the ciphertext vs the efficiency of the overall protocol.
%



\bibliographystyle{IEEEtran}
\bibliography{ref}
\end{document}